\documentclass[zamm,a4paper,fleqn]{w-art}
\usepackage{times}
\usepackage{w-thm}
\usepackage{cite}
\usepackage{amsmath,psfrag}
\usepackage{amssymb}
\usepackage[]{graphicx}
\begin{document}
\DOIsuffix{theDOIsuffix}
\Volume{VV}
\Issue{I}
\Month{MM}
\Year{YYYY}


\Receiveddate{XXXX}
\Reviseddate{XXXX}
\Accepteddate{XXXX}
\Dateposted{XXXX}



\title[Modeling the cyclic creep in the finite strain range ]{Modeling of cyclic creep in the finite strain range using a nested multiplicative split}


\author[A. V. Shutov]{Alexey V. Shutov\inst{1,2}%
  \footnote{Corresponding author,~e-mail:~\textsf{alexey.v.shutov@gmail.com},
            Phone: +07\,383\,333\,14 46,
            Fax: +07\,383\,333\,16 12}}
\address[\inst{1}]{Lavrentyev Institute of Hydrodynamics, pr. Lavrentyva 15, 630090, Novosibirsk, Russia}
\author[A. Yu. Larichkin]{Alexey Yu. Larichkin\inst{1,2,}}
\address[\inst{2}]{Novosibirsk State University, ul. Pirogova 2, 630090, Novosibirsk, Russia}
\author[V. A. Shutov]{Valeriy A. Shutov\inst{3,}}
\address[\inst{3}]{Novosibirsk State University of Architecture, Design and Arts, Krasny Prospekt 38, 630099, Novosibirsk, Russia}
\begin{abstract}

A new phenomenological model of cyclic creep is proposed which is suitable for applications
involving finite creep deformations
of the material. The model accounts for the the effect of the transient
increase of the creep strain rate upon the load reversal.
In order to extend the applicability range of the model, the
creep process is fully coupled to the classical Kachanov-Rabonov damage evolution.
As a result, the proposed model describes all the three stages of creep.
Large strain kinematics is described in a geometrically exact manner using
the assumption of a nested multiplicative split, originally proposed by Lion for
finite strain plasticity. The model is thermodynamically admissible, objective, and w-invariant.
Implicit time integration of the proposed evolution equations is discussed.
The corresponding numerical algorithm is implemented into the commercial FEM code MSC.MARC.
Using this code, the model is validated using real experimental data
on cyclic torsion of a thick-walled tubular specimen made of the D16T aluminium
alloy.
The numerically computed stress distribution exhibits a ``skeletal point" within the specimen.
\end{abstract}
\maketitle                   





\section{Introduction }

In numerous industrial applications, hight-temperature metallic structural components are loaded
by large stresses. Under such conditions
their long term behaviour depends on the creep-related effects, such
as the accumulation of irreversible creep deformations, redistribution of stresses, creep-induced anisotropy
and creep damage \cite{Boyle, AltenbachNaumenko}.
In the current study, the classical phenomenological approach to the modeling
of creep is developed within the framework of irreversible thermodynamics.

In the static case when applied stresses and temperature are constant, one usually identifies three stages
of creep: primary stage (characterized by transient or non-stationary response), secondary stage (steady-state creep) and tertiary stage (damage-dominated creep).
The creep rate typically reduces during the transition from the primary creep to the steady-state creep.
It is commonly accepted that this creep rate reduction is
caused by the material hardening, whereas the steady-state phase is characterized by a balance between hardening and recovery processes
\cite{Bailey}.
Apart from the primary creep under constant stress, the transient creep phenomenon occurs also
immediately after rapid changes of applied stresses. In particular,
a non-stationary creep is observed if the applied stresses are reversed. This transient process is characterized
on the macroscopic level by a relatively short period of an increased creep rate. During the holding time after the stress reversal, this transient process is
followed by the saturation to a reduced steady-state creep rate (cf., for example, Chapter 2.3 of Ref. \cite{AltenbachNaumenko}).
The accelerated creep strain rate after the stress reversal
($\sigma \mapsto -\sigma$) was experimentally observed in many studies (see, among others, \cite{Ohashi,Corum, Ohno1985}).
Another macroscopic manifestation of a non-stationary creep is as follows: After
an abrupt drop of the applied stress from $\sigma_1$ to $\sigma_2$ ($\sigma_1 > \sigma_2$)
the resulting creep rate is lower than the steady-state
creep rate observed in the material under $\sigma_2$.
At the same time, after an abrupt stress jump from $\sigma_2$ to $\sigma_1$, the resulting creep rate is higher than the
steady-state creep rate corresponding to the constant
stress level $\sigma_2$ \cite{Straub}.
The creep recovery upon
the removal of applied stresses \cite{AltenbachNaumenko} is another transient effect of this kind.

The above mentioned effects indicate that the creep is a loading-history
dependent anisotropic phenomenon.
An accurate description of the transient creep response of the material
is necessary for the correct analysis of high-temperature industrial components operating under cyclic
loads \cite{Gorash}.
The effect of the process-induced anisotropy is less evident in case
of a strain-controlled loading, but, nevertheless, it still has to be accounted
for (cf. \cite{Scholz2005, Naumenko2011, Kostenko2013}).
 Let us briefly discuss the main phenomenological approaches to cyclic creep.
The most simple macroscopic description of the transient stage is provided by the concept of isotropic
creep-hardening \cite{Boyle}\footnote{
Even more simple transient creep models can be constructed if the material parameters are assumed as explicit functions of time. Such
explicit time hardening models are not considered in the current study, since they violate the objectivity principle:
Explicit time hardening violates the invariance of constitutive equations with respect to the change of the time reference point.
For the same reason, pseudo-plasticity models which are based on pseudo-stresses \cite{Faruque} are not considered here.}.
Unfortunately, as one may expect, the simple isotropy assumption is not sufficient in many practical cases \cite{Findley1979, MurakamiOhno, Ohashi}.
One of the most popular macroscopic approaches to the creep-induced anisotropy is based on the concept
of kinematic hardening, which was borrowed from the phenomenological plasticity.
Within this concept, backstresses are introduced as a measure of the accumulated anisotropy
\footnote{The pioneering paper on the simulation of the kinematic hardening using
 backstresses was published by Prager in 1935 \cite{Prager1935}.}.
Although the concept of backstresses is nowadays wide spread, there are
 different micromechanical
interpretations in case of polycristalline materials.
One may assume that some backstresses represent a resistance
to dislocation motion caused by pile-ups of dislocations \cite{GarmestaniAUnified, Zolochevsky} (atomistic scale),
while other backstresses represent an internal residual stress field caused by
plastic strain incompatibilities between grains (mesoscopic scale) \cite{Feugas1999}. In either way,
backstresses superimpose with the applied mechanical stresses such that the
creep is governed by the effective stress, computed according to the formula
$$\text{effective stress} = \text{mechanical stress} - \text{backstress}.$$
Following this concept, the anisotropic creep is assumed to be governed by the effective
stresses \cite{MalininKhadjinsky, LemaitreChaboche, Kawai1995, Klebanov}.
In particular, the assumption is made that the creep strain rate is coaxial to the deviatoric effective stresses,
not to the deviatoric part of the stresses itself \cite{FaruqueZaman}.

The evolution of the backstress $X$ can be described in a hardening/recovery format (see Section 2), where
the hardening term represents the microstructural changes associated with the material strengthening
and the recovery term is typically related to the softening
of the material
$$ \dot{X} = \dot{X}|_{\text{hardening}} - \dot{X}|_{\text{recovery}}.$$
The hardening term is usually strain controlled; therefore it is assumed as a homogeneous function
of the creep strain rate $\dot{\varepsilon}_{\text{cr}}$
$$ \dot{X}|_{\text{hardening}} (\alpha \dot{\varepsilon}_{\text{cr}}) =
\alpha \dot{X}|_{\text{hardening}} (\dot{\varepsilon}_{\text{cr}}) \quad \text{for all} \ \alpha \geq 0.$$
Dynamic recovery of the backstress implies strain-controlled creep hardening (e.g. Armstrong-Frederick-like behaviour, or,
equivalently, endochronic Maxwell-like behaviour)
$$\dot{X}|_{\text{recovery}} = \| \dot{\varepsilon}_{\text{cr}} \|  {F}(X),$$
where $F(X)$ is a suitable function of the backstress $X$ and $\| \dot{\varepsilon}_{\text{cr}} \|$ is a norm of the creep strain rate.
The dynamic recovery
format was used, among others, in \cite{Naumenko2010, Kostenko2013}. This type of material behaviour is also known as a strain-activated recovery.
Alternatively, one may assume the so-called
static recovery, which
implies time-controlled evolution (e.g. Maxwell-like behaviour)
$$\dot{X}|_{\text{recovery}} = \dot{X}|_{\text{recovery}} (X).$$
Since this process is partially driven by the diffusion, it is highly temperature dependent; for that reason
static recovery is also known as a temperature-controlled recovery.
The static recovery format was implemented in \cite{MalininKhadjinsky, Slavik, Krieg, Kawai1995, FaruqueZaman, Zolochevsky}.  
Backstress-based creep models combining both static and dynamic recovery are also known (see eq. (8) in \cite{Mroz}, eq. (62) in \cite{Kawai1995}
or eq.  (3) in \cite{Velay2006}). Such a combined static/dynamic approach is utilized in the current study as well.

The basic hardening/recovery format mentioned above can be further specified in a number of ways.
The evolution equations proposed in \cite{Krieg} are motivated by microstructural information, namely, by experimental measurements of the
(average) dislocation cell size.
An additional constitutive assumption was used in  \cite{FaruqueZaman} to capture the history-dependent material response:
the stress response is assumed to depend on the maximum value of the backstress, achieved in the previous history.
In some studies, the saturation level for backstresses depends on the applied stresses \cite{MalininKhadjinsky, FaruqueZaman, Gorash}.
In \cite{Zolochevsky, Zolochevsky2005} the backstresses are not necessarily deviatoric;
the hydrostatic component of the backstrasses is relevant
for pressure-sensitive materials.

The safety analysis of industrial components is mostly concerned with small strain creep.
Comprehensive reviews of different creep hardening rules in the small strain context can be found in
\cite{Chaboche, Ohno1990}.
\footnote{A small strain creep analysis with finite deflections and rotations of
thin-walled structures (von Karman's approximation) was discussed in \cite{Altenbach1997}.}
At the same time, analysis of accident scenarios may involve finite strains as well;
moreover, a number of metal forming applications
involve creep processes in the finite strain range.
Unfortunately, there is only a few publications devoted to the finite strain
creep analysis (see, among others, \cite{Billington, Krieg}). The aim of the current paper is to fill this gap; we apply
the state of the art methodology of anisotropic multiplicative
plasticity to the specific problems of creep mechanics.
Here we choose the multiplicative framework since it has numerous advantages
over alternative approaches \cite{ShutovAnalysisOfSome}.
Following \cite{Lion}, the classical decomposition of the deformation gradient
into inelastic (creep) and elastic parts is now supplemented by a nested multiplicative split of the
inelastic part into some dissipative and conservative parts.
This decomposition allows one to incorporate the nonlinear
kinematic hardening in a thermodynamically consistent way
(see \cite{Helm1, ShutovKreissig2008, Vladimirov, ShutovIncrem, Broecker, Kiesling}
among others).

The paper is organized as follows.
In Section 2 the constitutive equations
of the developed creep model are presented.
In Section 3 the numerical implementation of the proposed model
is discussed.
In Section 4 the model is validated using the experimental data on the torsion of a thick-walled tubular sample made of D16T aluminum alloy.

A coordinate-free tensor formalism (direct tensor notation) is used in the current study.
Second- and fourth-rank tensors in $\mathbb{R}^3$ are denoted by bold symbols. Notations $\text{tr}( \cdot )$,
$( \cdot )^{\text{D}}$, $( \cdot )^{\text{T}}$, $( \cdot )^{\text{-T}}$,  $\det( \cdot )$ stand for the trace, deviatoric part, transposition,
 inverse of transposed, and determinant, respectively. The symmetric part, scalar product of two second-rank tensors (double contraction), the Frobenius norm, and the
unimodular part are defined as follows
\begin{equation}\label{Notations}
\text{sym}( \mathbf{A} ) := \frac{1}{2} ( \mathbf{A} + \mathbf{A}^{\text{T}}), \quad
\mathbf{A} : \mathbf{B} \ := \text{tr} (\mathbf{A} \ \mathbf{B}^{\text{T}}), \quad
\| \mathbf{A} \| := \sqrt{ \mathbf{A}:\mathbf{A} }, \quad
\overline{\mathbf{A}} := (\det(\mathbf{A}))^{-1/3} \ \mathbf{A}.
\end{equation}
Suffixes
$(\cdot)_{\text{el}}$, $(\cdot)_{\text{cr}}$,
$(\cdot)_{\text{ii}}$, and $(\cdot)_{\text{ie}}$
stand for ``elastic", ``creep", ``inelastic-inelastic", ``inelastic-elastic".
Since the presentation is coordinate free, these suffixes can not be
mistaken for tensor coordinates.

\section{Material model of anisotropic creep with isotropic damage}

The constitutive equations of creep will be combined with the classical Kachanov-Rabotnov approach to creep damage.
Toward that end we introduce the Rabotnov damage variable $\omega \in [0,1]$.
Schematically, $\omega = 0$ corresponds to the intact material and
$\omega = 1$ characterizes a fully destroyed material with a macroscopic crack \cite{Kachanov, Rabotnov, MileikoRabotnov}.

\subsection{Small strain case}

\begin{figure}\centering
\psfrag{A}[m][][1][0]{$\boldsymbol{\varepsilon}_{\text{cr}}$}
\psfrag{B}[m][][1][0]{$\boldsymbol{\varepsilon}_{\text{e}}$}
\psfrag{C}[m][][1][0]{$\boldsymbol{\varepsilon}$}
\psfrag{D}[m][][1][0]{substructure}
\psfrag{E}[m][][1][0]{$\boldsymbol{\varepsilon}_{\text{ii}}$}
\psfrag{F}[m][][1][0]{$\boldsymbol{\varepsilon}_{\text{ie}}$}
\psfrag{Q}[m][][1][0]{a)}
\psfrag{R}[m][][1][0]{b)}
\psfrag{G}[m][][1][0]{$\varkappa_{\text{stat}}$}
\psfrag{M}[m][][1][0]{$\varkappa_{\text{dynam}}$}
\psfrag{H}[m][][1][0]{$c$}
\psfrag{K}[m][][1][0]{$\alpha^{\lambda}_1, \alpha^{\lambda}_2, \alpha,...$}
\psfrag{L}[m][][1][0]{$k, \mu$}
\scalebox{0.95}{\includegraphics{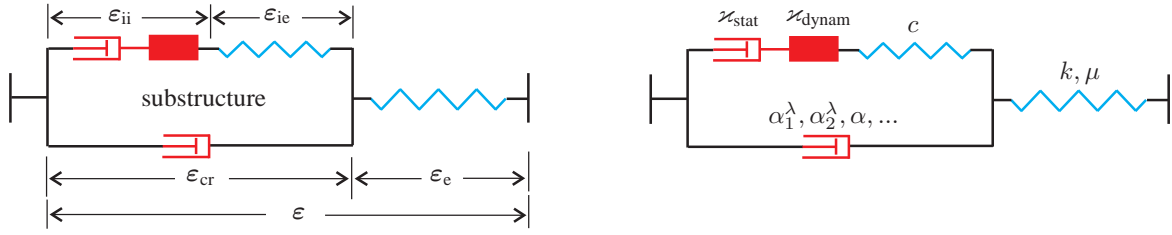}}
\caption{Rheological model of creep with nonlinear kinematic hardening, which
includes static and dynamic recovery: decomposition of the infinitesimal strain  (left) and introduced material parameters (right). \label{fig1}}
\end{figure}

We start with a small strain version of the material model, which is extremely simple since
geometric nonlinearities are neglected.
To visualize the main modeling assumptions we employ the
 rheological interpretation shown in Fig. \ref{fig1}(left).
The rheological model consists of two generalized Maxwell
bodies (accounting for static and dynamic recovery), a Hooke body and a modified Newton body.
The overall infinitesimal strain tensor $\boldsymbol{\varepsilon}$ is decomposed additively
into the creep strain $\boldsymbol{\varepsilon}_{\text{cr}}$
and the elastic strain $\boldsymbol{\varepsilon}_{\text{e}}$. The
creep strain itself is decomposed into the
dissipative part
$\boldsymbol{\varepsilon}_{\text{ii}}$ and
the conservative part
$\boldsymbol{\varepsilon}_{\text{ie}}$
\begin{equation}\label{AddDecomposit}
\boldsymbol{\varepsilon} = \boldsymbol{\varepsilon}_{\text{cr}} + \boldsymbol{\varepsilon}_{\text{e}},
\quad
\boldsymbol{\varepsilon}_{\text{cr}} = \boldsymbol{\varepsilon}_{\text{ii}} + \boldsymbol{\varepsilon}_{\text{ie}}.
\end{equation}
Using these strain variables we define the Helmholz free energy per unit mass
\begin{equation}\label{HelmholzSmalStr}
\psi = \psi(\boldsymbol{\varepsilon}_{\text{e}},\boldsymbol{\varepsilon}_{\text{ie}},\omega) = \psi_{\text{el}}(\boldsymbol{\varepsilon}_{\text{e}}, \omega)
+ \psi_{\text{kin}}(\boldsymbol{\varepsilon}_{\text{ie}},\omega),
\end{equation}
where $\psi_{\text{el}}(\boldsymbol{\varepsilon}_{\text{e}}, \omega)$ is the energy storage due to the macroscopic elastic deformations
of the crystal lattice and $\psi_{\text{kin}}(\boldsymbol{\varepsilon}_{\text{ie}},\omega)$ is the part of the energy stored in the defects
of the crystal structure, associated with the kinematic hardening.\footnote{This additive split can be motivated by the
rheological model shown in Fig. \ref{fig1}a.}
The stress tensor $\boldsymbol{\sigma}$ and the backstress tensor $\boldsymbol{x}$ are then computed through
\begin{equation}\label{HelmholzStresses}
\boldsymbol{\sigma} =  \rho
\frac{\displaystyle \partial \psi_{\text{el}}(\boldsymbol{\varepsilon}_{\text{e}}, \omega)}{\displaystyle \partial \boldsymbol{\varepsilon}_{\text{e}}}, \quad
\boldsymbol{x} =  \rho \frac{\displaystyle \partial
\psi_{\text{kin}}(\boldsymbol{\varepsilon}_{\text{ie}},\omega)}
{ \displaystyle \partial \boldsymbol{\varepsilon}_{\text{ie}}},
\end{equation}
where $\rho$ is the mass density.
The formulation of the model may employ various assumptions governing the free energy; however,
to be definite, we use here the following quadratic strain energy function
\begin{equation}\label{QnadraticStrainEnergy}
\rho \psi_{\text{el}} (\boldsymbol{\varepsilon}_{\text{e}}, \omega ) =
(1-\omega) \frac{k}{2} (\text{tr} \boldsymbol{\varepsilon}_{\text{e}})^2 + (1-\omega) \mu \boldsymbol{\varepsilon}_{\text{e}}^{\text{D}} :
\boldsymbol{\varepsilon}_{\text{e}}^{\text{D}}, \quad
\rho \psi_{\text{kin}} (\boldsymbol{\varepsilon}_{\text{ie}}, \omega ) =
(1-\omega) \frac{c}{2} \boldsymbol{\varepsilon}_{\text{ie}}^{\text{D}} : \boldsymbol{\varepsilon}_{\text{ie}}^{\text{D}},
\end{equation}
where $k$, $\mu$, and $c$ are material parameters characterizing
the intact material. Substituting this into \eqref{HelmholzStresses} we arrive at
\begin{equation}\label{Stresses}
\boldsymbol{\sigma} = k \ (1-\omega) \ \text{tr}(\boldsymbol{\varepsilon}_{\text{e}}) \boldsymbol{1} + 2 \mu (1-\omega) \ \boldsymbol{\varepsilon}_{\text{e}}^{\text{D}}, \quad
\boldsymbol{x} = c (1-\omega) \ \boldsymbol{\varepsilon}_{\text{ie}}^{\text{D}},
\end{equation}
where $\boldsymbol{1}$ is the second-rank identity tensor.
According to \eqref{Stresses}, both the macroscopic elastic properties (bulk and shear moduli $k$ and $\mu$)
and the elastic properties of the substructure (shear modulus of substructure $c$) deteriorate with
damage.\footnote{For simplicity we assume here that
the bulk modulus and the shear modulus deteriorate with the same rate. In a more general case
one may introduce two different
rates \cite{ShutovDuctDamage} or even two different damage variables \cite{Zapara2010, Zapara2012}.}
The effective stress is defined through
\begin{equation}\label{EffStress}
\boldsymbol{\sigma}_{\text{eff}} := \boldsymbol{\sigma} - \boldsymbol{x}.
\end{equation}

We assume that the effective stress $\boldsymbol{\sigma}_{\text{eff}}$
is the driving force of the global creep process.
The framework which will be developed in the current study
allows the creep strain rate to be an arbitrary isotropic function of the
effective stress.
However, for simplicity, we will restrict our attention to incompressible creep flow.
In order to be more specific, we need the following preparations.
First, let $\mathbf{A}$ be an arbitrary symmetric second-rank tensor.
Its \emph{regularized} maximum eigenvalue $\sigma_{\text{reg max}}(\mathbf{A})$ is defined through the formula
\begin{equation}\label{SigMaxReg}
\sigma_{\text{reg max}}(\mathbf{A}) := ( \langle a_1 \rangle^R + \langle a_2 \rangle^R + \langle a_3 \rangle^R)^{1/R},
\quad \{ a_1, a_2, a_3 \} := \text{eigenvalues of} \ \mathbf{A},
\end{equation}
where $\langle x \rangle := \max(0,x)$; $R>1$ is a
regularization parameter.\footnote{The maximum positive eigenvalue is obtained as $R \rightarrow \infty$.
By the Davis-Lewis theorem formulated for spectral functions (cf. \cite{Borwein}),
$\sigma_{\text{reg max}}(\mathbf{A})$ is a convex function of
$\mathbf{A}$ for $R \geq 1$.
We need the maximum positive eigenvalue since the creep-related material properties are commonly
assumed to depend on this quantity \cite{AltenbachNaumenko, AltenbachAltenbachZolochevsky}.
In this study, however, we are using its regularized counterpart \eqref{SigMaxReg} to ensure
that the derivative of $\sigma_{\text{reg max}} (\boldsymbol{\sigma}_{\text{eff}})$
with respect to the effective stress tensor $\boldsymbol{\sigma}_{\text{eff}}$ is a continuous function.}
Next, two stress invariants (equivalent stresses)
are defined in the following way
\begin{equation}\label{equivStresslamb}
\sigma^{\lambda}_{\text{eq}} := \alpha^{\lambda}_1 \sigma_{\text{reg max}} (\boldsymbol{\sigma}_{\text{eff}}) +
\alpha^{\lambda}_2 \ \sqrt{\frac{3}{2}} \ \| \boldsymbol{\sigma}^{\text{D}}_{\text{eff}}  \| +
(1 - \alpha^{\lambda}_1 - \alpha^{\lambda}_2) \ \text{tr}(\boldsymbol{\sigma}_{\text{eff}}),
\end{equation}
\begin{equation}\label{equivStressDire}
\sigma_{\text{eq}} := \alpha \ \frac{3}{2} \ \sigma_{\text{reg max}}(\boldsymbol{\sigma}^{\text{D}}_{\text{eff}})  +
(1-\alpha) \ \sqrt{\frac{3}{2}} \ \| \boldsymbol{\sigma}^{\text{D}}_{\text{eff}}  \|,
\end{equation}
where $\alpha^{\lambda}_1, \alpha^{\lambda}_2, \alpha \geq 0$ are constant weighting factors.
Note that these function are homogeneous functions of $\boldsymbol{\sigma}_{\text{eff}}$ of degree one.\footnote{
These invariants are called ``equivalent stresses" since $\sigma^{\lambda}_{\text{eq}} = \sigma_{\text{eq}} = \sigma$ for
$\boldsymbol{\sigma}_{\text{eff}} = \sigma \ \mathbf{n} \otimes \mathbf{n}$ with $ \| \mathbf{n} \| =1$ and $\sigma > 0$.}
Using these, we postulate the flow rule in the form
\begin{equation}\label{CreepEvolut}
\dot{\boldsymbol{\varepsilon}}_{\text{cr}} = \lambda (\sigma^{\lambda}_{\text{eq}},\omega)
\frac{\partial \sigma_{\text{eq}}}{\partial \boldsymbol{\sigma}_{\text{eff}}},
\end{equation}
where $\lambda
(\sigma^{\lambda}_{\text{eq}}, \omega)$ is a suitable function of the equivalent stress and damage $\omega$.

\textbf{Remark 1.}
According to \eqref{CreepEvolut}, the multiplier $\lambda (\sigma^{\lambda}_{\text{eq}},\omega)$
 controls the intensity of the creep flow and $\frac{\displaystyle \partial
\sigma_{\text{eq}}}{\displaystyle \partial \boldsymbol{\sigma}_{\text{eff}}}$
gives its direction. In fact, the flow rule \eqref{CreepEvolut} is based on the
assumption that there is a creep flow potential. This assumption
is quite common in the creep mechanics, even dealing with
anisotropic materials \cite{SosninCreepPotential}. $\Box$

Following the classical Kachanov-Rabotnov approach we postulate
\begin{equation}\label{KachanRabot}
\lambda(\sigma^{\lambda}_{\text{eq}},\omega) = (1-\omega)^{-m} \ \lambda_{\text{undamaged}} (\sigma^{\lambda}_{\text{eq}}),
\end{equation}
where $m$ is a material parameter and $\lambda_{\text{undamaged}} (\sigma^{\lambda}_{\text{eq}})$
is the creep strain rate of the undamaged material.
In general, any non-negative and smooth function $\lambda_{\text{undamaged}} (\sigma^{\lambda}_{\text{eq}})$ can be used if the
natural restriction
$\lambda_{\text{undamaged}}|_{\sigma^{\lambda}_{\text{eq}} =0} = 0$ is
satisifed. The following monotonic functions of $\sigma^{\lambda}_{\text{eq}}$ are
frequently used in the phenomenological creep modeling (see, for example, \cite{AltenbachNaumenko})
\begin{equation}\label{CreepRateNorton}
\lambda_{\text{undamaged}} (\sigma^{\lambda}_{\text{eq}}) = A ( \sigma^{\lambda}_{\text{eq}}/ f_0 )^n, \quad f_0:=1 \text{MPa},
\end{equation}
\begin{equation}\label{CreepRateSoderberg}
\lambda_{\text{undamaged}} (\sigma^{\lambda}_{\text{eq}}) = A ( \exp(\sigma^{\lambda}_{\text{eq}}/\sigma_0) - 1 ),
\end{equation}
\begin{equation}\label{CreepRatePrandtl}
\lambda_{\text{undamaged}} (\sigma^{\lambda}_{\text{eq}}) = A \sinh(\sigma^{\lambda}_{\text{eq}}/\sigma_0),
\end{equation}
\begin{equation}\label{CreepRateJohnson}
\lambda_{\text{undamaged}} (\sigma^{\lambda}_{\text{eq}}) = A_1 ( \sigma^{\lambda}_{\text{eq}}/ f_0 )^{n_1} +  A_2 ( \sigma^{\lambda}_{\text{eq}} / f_0 )^{n_2},
  \quad f_0:=1 \text{MPa},
\end{equation}
\begin{equation}\label{CreepRateGarofalo}
\lambda_{\text{undamaged}} (\sigma^{\lambda}_{\text{eq}}) = A \big( \sinh(\sigma^{\lambda}_{\text{eq}}/\sigma_0) \big)^n,
\end{equation}
where $A >0$, $A_1 >0$, $A_2 >0$, $\sigma_0 >0$, $n \geq 1$, $n_1 \geq 1$, $n_2 \geq 1$ are material parameters.

Next, we assume that the backstress $\mathbf{x}$ is the driving force
for the dissipative processes on the substructural level. In particular,
the saturation of the kinematic hardening is described using the following flow rule
\begin{equation}\label{KinematHardRecov}
\dot{\boldsymbol{\varepsilon}}_{\text{ii}} = \varkappa_{\text{dynam}} \
 \|\dot{\boldsymbol{\varepsilon}}_{\text{cr}}\|  \ \mathbf{x}^{\text{D}} + \varkappa_{\text{stat}} \ \mathbf{x}^{\text{D}},
\end{equation}
where $\varkappa_{\text{dynam}} \geq 0$ and $\varkappa_{\text{stat}} \geq 0$
are the parameters of dynamic and static recovery.
Note that \eqref{KinematHardRecov} and the corresponding material parameters
are not influenced by damage.
This is a strong simplifying assumption, but, as will be clear from the model validation (cf. Section 4),
it yields good results.

\textbf{Remark 2.}
The dynamic recovery term $\varkappa_{\text{dynam}} \
 \|\dot{\boldsymbol{\varepsilon}}_{\text{cr}}\|  \ \mathbf{x}$
 is needed to capture the following important effect, observed in
 experiments on real materials: the higher the applied stress is, the
 shorter the transient stage \cite{MileikoRabotnov}.
 Indeed, for high applied stress the creep rate is high thus leading to the fast saturation of $\mathbf{x}$.
 On the other hand, the dynamic term alone may be not enough to
 obtain a plausible mechanical response. In particular, the creep models
 with kinematic hardening without static recovery are prone to the following unphysical behaviour
 under static loading conditions:
 If the applied stress $\sigma$ is smaller than the saturation level for the backstress $\mathbf{x}$,
 then after a certain holding time the backstress equilibrates the applied stress and the
 creep rate becomes exactly zero. In order to prevent such unrealistic behaviour, the static recovery
 term $\varkappa_{\text{stat}} \ \mathbf{x}$ is introduced in \eqref{KinematHardRecov}.
 One important implication of the static recovery is that the deformation-induced backstresses
relax even if the creep strain is frozen: $\mathbf{x} \rightarrow \mathbf{0}$ as
$t \rightarrow \infty$ and $\dot{\boldsymbol{\varepsilon}}_{\text{cr}} = \mathbf{0}$.  $\Box$

\textbf{Remark 3.} In this study we assume for simplicity $\varkappa_{\text{dynam}} = const$,
which is sufficient for our goals.
However, in some materials
the saturation level of the backstresses is nearly proportional to the applied stresses. In order
to capture this effect, one may consider $\varkappa$ to be a (positive) function of the applied stress $\boldsymbol \sigma$. $\Box$

Analogously to \eqref{equivStresslamb}, to render the evolution of the
damage variable $\omega$ we introduce a damage-related equivalent stress
\begin{equation}\label{equivStressOmeg}
\sigma^{\omega}_{\text{eq}} := \alpha^{\omega}_1 \sigma_{\text{reg max}} (\boldsymbol{\sigma}_{\text{eff}}) +
\alpha^{\omega}_2 \ \sqrt{\frac{3}{2}} \  \| \boldsymbol{\sigma}^{\text{D}}_{\text{eff}}  \| +
(1 - \alpha^{\omega}_1 - \alpha^{\omega}_2) \ \text{tr}(\boldsymbol{\sigma}_{\text{eff}}), \quad
\end{equation}
where $\alpha^{\omega}_1 \geq 0$ and $\alpha^{\omega}_2 \geq 0$ are material constants.
Then the damage evolution is given by the classical Kachanov-Rabotnov relation
\begin{equation}\label{equivStressOmegEv}
\dot{\omega} = B (1-\omega)^{-l} (\sigma^{\omega}_{\text{eq}})^k,
\end{equation}
where $B$, $k$, and $l$ are material parameters.

Finally, in order to close the system of constitutive equations
we put the following initial conditions
\begin{equation}\label{InitCond}
\boldsymbol{\varepsilon}_{\text{cr}}|_{t=0} = \boldsymbol{\varepsilon}^0_{\text{cr}}, \quad
\boldsymbol{\varepsilon}_{\text{ii}}|_{t=0} = \boldsymbol{\varepsilon}^0_{\text{ii}}, \quad
\omega|_{t=0} = \omega^0.
\end{equation}
In general, the initial values $\boldsymbol{\varepsilon}^0_{\text{cr}}$ and $\boldsymbol{\varepsilon}^0_{\text{ii}}$
can be seen as additional material parameters, which characterize the material at $t=0$.
Observe that the introduced material parameters
can be interpreted in terms of the rheological model, as shown in Fig. \ref{fig1}(right).

\subsection{Generalization to finite strains}

\begin{figure}\centering
\psfrag{A}[m][][1][0]{${\mathbf F}_{\text{cr}}$}
\psfrag{B}[m][][1][0]{${\mathbf F}_{\text{e}}$}
\psfrag{C}[m][][1][0]{${\mathbf F}$}
\psfrag{D}[m][][1][0]{substructure}
\psfrag{E}[m][][1][0]{${\mathbf F}_{\text{ii}}$}
\psfrag{F}[m][][1][0]{${\mathbf F}_{\text{ie}}$}
\psfrag{Q}[m][][1][0]{a)}
\psfrag{R}[m][][1][0]{b)}
\psfrag{I}[m][][1][0]{$\tilde{\mathcal{K}}$}
\psfrag{J}[m][][1][0]{$\check{\mathcal{K}}$}
\psfrag{K}[m][][1][0]{$\hat{\mathcal{K}}$}
\psfrag{S}[m][][1][0]{$\mathcal{K}$}
\psfrag{L}[m][][1][0]{$\mathbf F_{\text{ii}}$}
\psfrag{M}[m][][1][0]{$\mathbf F_{\text{cr}}$}
\psfrag{N}[m][][1][0]{$\check{\mathbf F}_{\text{ie}}$}
\psfrag{O}[m][][1][0]{$\hat{\mathbf F}_{\text{e}}$}
\psfrag{P}[m][][1][0]{$\mathbf F$}
\scalebox{0.95}{\includegraphics{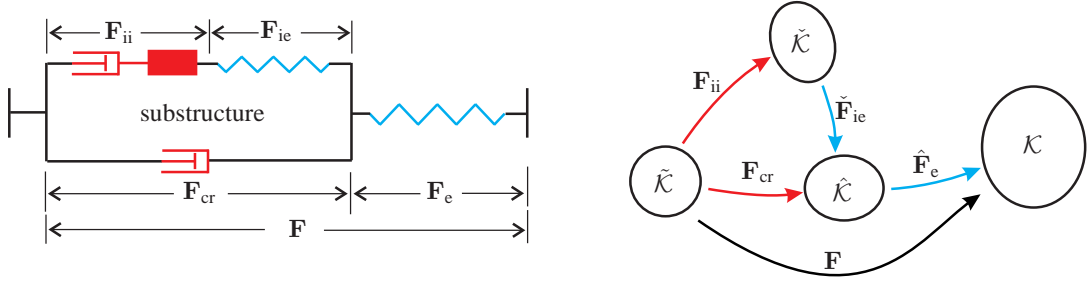}}
\caption{Rheological model of creep with nonlinear kinematic hardening, which
includes static and dynamic recovery: partial deformations  (left) commutative diagram pertaining to the nested multiplicative split
of the deformation gradient (right). \label{fig2}}
\end{figure}

\textbf{Description of kinematics.} In this subsection we generalize the previously presented material model to the finite strain range by
utilizing a nested multiplicative split originally proposed by Lion in \cite{Lion}. This nested split is essentially
motivated by the rheological model show in Fig. \ref{fig2}(left).\footnote{The use of other rheological models
in the finite strain creep was already discussed in \cite{Krawietz, Palmov}.}
Let $\mathbf F$ be the deformation gradient mapping the local reference configuration $\widetilde{\mathcal{K}}$
to the current configuration $\mathcal{K}$. The deformation gradient $\mathbf F$
is decomposed multiplicatively into the creep part $\mathbf F_{\text{cr}}$ and
the elastic part $\mathbf F_{\text{e}}$; the creep part itself is decomposed into the
dissipative part $\mathbf F_{\text{ii}}$ and a conservative part $\mathbf F_{\text{ie}}$
\begin{equation}\label{mude}
\mathbf F = \hat{\mathbf F}_{\text{e}} \mathbf F_{\text{cr}}, \quad
\mathbf F_{\text{cr}} = \check{\mathbf F}_{\text{ie}} \mathbf F_{\text{ii}}.
\end{equation}
These multiplicative decompositions can be seen as a generalization of the additive split \eqref{AddDecomposit};
they are summarized in a commutative diagram shown in Fig. \ref{fig2}(right).
The first decomposition defines the stress-free configuration $\hat{\mathcal{K}}$
and the second decomposition implies the configuration of kinematic hardening $\check{\mathcal{K}}$.
Next, we introduce the right Cauchy-Green tensor $\mathbf C$,
the right Cauchy-Green tensor of creep $\mathbf C_{\text{cr}}$, and the
inelastic right Cauchy-Green tensor of substructure $\mathbf C_{\text{ii}}$
\begin{equation}\label{RCG}
\mathbf C := \mathbf F ^{\text{T}} \mathbf F, \quad
\mathbf C_{\text{cr}} := \mathbf F_{\text{cr}}^{\text{T}} \mathbf F_{\text{cr}}, \quad
\mathbf C_{\text{ii}} := \mathbf F_{\text{ii}}^{\text{T}} \mathbf F_{\text{ii}}.
\end{equation}
Note that these tensors operate on the reference configuration.
Further, we introduce the elastic right Cauchy-Green tensor $\hat{\mathbf C}_{\text{e}}$
operating on the intermediate configuration $\hat{\mathcal{K}}$
and the elastic right Cauchy-Green tensor of substructure $\check{\mathbf C}_{\text{ie}}$, which operates on the
configuration of the kinematic hardening $\check{\mathcal{K}}$
\begin{equation}\label{RCG_elas}
\hat{\mathbf C}_{\text{e}} := \mathbf F_{\text{e}}^{\text{T}} \mathbf F_{\text{e}}, \quad
\check{\mathbf C}_{\text{ie}} := \mathbf F_{\text{ie}}^{\text{T}} \mathbf F_{\text{ie}}.
\end{equation}
The rate of the creep flow is captured with the gradient of the creep velocity $\mathbf L_{\text{cr}}$
and the creep strain rate $\mathbf D_{\text{cr}}$, both operating on $\hat{\mathcal{K}}$
\begin{equation}\label{CreepRate}
\hat{\mathbf L}_{\text{cr}} := \dot{\mathbf F}_{\text{cr}} \mathbf F^{-1}_{\text{cr}}, \quad
\hat{\mathbf D}_{\text{cr}} := \text{sym} (\hat{\mathbf L}_{\text{cr}}).
\end{equation}
Analogously, the inelastic strain rate $\mathbf D_{\text{ii}}$ on the substructural level, which operates on
$\check{\mathcal{K}}$,  is defined as follows
\begin{equation}\label{InelasSubstr}
\check{\mathbf L}_{\text{ii}} := \dot{\mathbf F}_{\text{ii}} \mathbf F^{-1}_{\text{ii}}, \quad
\check{\mathbf D}_{\text{ii}} := \text{sym} (\check{\mathbf L}_{\text{ii}}).
\end{equation}

\textbf{Free energy and stresses.} Aiming at a thermodynamically consistent formulation, we introduce the following ansatz for the free energy density
per unit mass
(cf. \cite{Lion, Helm1, ShutovKreissig2008})
\begin{equation}\label{freeen}
\psi=\psi(\hat{\mathbf C}_{\text{e}}, \check{\mathbf C}_{\text{ie}}, \omega)=
\psi_{\text{el}} (\hat{\mathbf C}_{\text{e}}, \omega) + \psi_{\text{kin}} (\check{\mathbf C}_{\text{ie}}, \omega).
\end{equation}
Its microstructural interpretation is the same as for \eqref{HelmholzSmalStr}.
Just as in the small-strain case, this additive split can be motivated by
the rheological model shown on Fig. \ref{fig2}(left).
The framework proposed in the current study is valid for
arbitrary isotropic functions $\psi_{\text{el}}$ and $\psi_{\text{kin}}$. However, to be definite, we use the neo-Hooke-like
assumptions for the energy storage
\begin{equation}\label{neoHooke1}
\rho_{\scriptscriptstyle \text{R}} \psi_{\text{el}} (\hat{\mathbf C}_{\text{e}}, \omega) = (1-\omega)
\frac{k}{50}\big( (\text{det} \mathbf{C}_{\text{e}})^{5/2} + (\text{det} \mathbf{C}_{\text{e}} )^{-5/2} -2 \big) + (1-\omega)
\frac{\mu}{2} \big( \text{tr} \ \overline{\hat{\mathbf{C}}_{\text{e}}} - 3 \big),
\end{equation}
\begin{equation}\label{neoHooke2}
\rho_{\scriptscriptstyle \text{R}}
\psi_{\text{kin}} (\check{\mathbf C}_{\text{ie}}, \omega) = (1-\omega)
\frac{c}{4}\big( \text{tr} \ \overline{\check{\mathbf{C}}_{\text{ie}}} - 3 \big).
\end{equation}
Here, $k >0$, $\mu > 0$, and $c \geq 0$ are the material constants already introduced in the small strain case;
$\rho_{\scriptscriptstyle \text{R}}$ stands for the mass density in the reference configuration.
The term $\frac{k}{50}\big( (\text{det} \mathbf{C}_{\text{e}})^{5/2} + (\text{det} \mathbf{C}_{\text{e}} )^{-5/2} -2 \big)$ which appears
on the right-hand side of \eqref{neoHooke1} corresponds to the volumetric part of the free energy. This special ansatz was proposed in \cite{HartmannNeff};
it is advantageous over many alternative assumptions since it implies that the free energy is a convex function of $\text{det} \mathbf{C}_{\text{e}}$.

Let $\mathbf T$ be the Cauchy stress tensor (true stresses).
The Kirchhoff stress tensor $\mathbf S$ (weighted Cauchy tensor), the second
Piola-Kirchhoff stress $\tilde{\mathbf T}$ operating on the reference configuration $\widetilde{\mathcal{K}}$, and the Kirchhoff stress $\hat{\mathbf S}$ operating
 on the stress-free configuration $\hat{\mathcal{K}}$ are defined through
\begin{equation}\label{Kirch}
\mathbf S : = (\text{det} \mathbf F) \mathbf T, \quad
\tilde{\mathbf T} := \mathbf F^{-1} \ \mathbf S \ \mathbf F^{-\text{T}}, \quad
\hat{\mathbf S} : = \mathbf F_{\text{e}}^{-1} \ \mathbf S \ \mathbf F_{\text{e}}^{-\text{T}}.
\end{equation}
Further, let $\check{\mathbf X}$ be a backstress tensor, operating on the configuration of kinematic hardening $\check{\mathcal{K}}$.
In the following we will interpret $\check{\mathbf X}$ as a generalized stress measure conjugate to the deformation rate $\check{\mathbf D}_{\text{ii}}$.
Its counterparts operating on the reference $\tilde{\mathcal{K}}$ and
on the stress-free configuration $\hat{\mathcal{K}}$ are obtained by the following pull-back and push-forward, respectively
\begin{equation}\label{Backstress}
\tilde{\mathbf X} := \mathbf F^{-1}_{\text{ii}} \  \check{\mathbf X} \ \mathbf F_{\text{ii}}^{-\text{T}}, \quad
\hat{\mathbf X} := \mathbf F_{\text{ie}} \  \check{\mathbf X} \ \mathbf F_{\text{ie}}^{\text{T}}.
\end{equation}
Additional details on the derivation of these generalized stresses can be found in \cite{ShutovKreissig2008}.
Now we postulate the following relations of hyperelastic type (cf. \cite{Lion, Helm1, ShutovKreissig2008})
\begin{equation}\label{Hyperelas}
\hat{\mathbf S} = 2 \ \rho_{\scriptscriptstyle \text{R}}
\frac{\displaystyle \partial \psi_{\text{el}}(\hat{\mathbf{C}}_{\text{e}}, \omega)}{\displaystyle \partial \hat{\mathbf{C}}_{\text{e}}}, \quad
\check{\mathbf X} = 2 \ \rho_{\scriptscriptstyle \text{R}}
\frac{\displaystyle \partial
\psi_{\text{kin}}(\check{\mathbf{C}}_{\text{ie}}, \omega)}
{\displaystyle \partial \check{\mathbf{C}}_{\text{ie}}}.
\end{equation}

\textbf{Evolution equations and thermodynamic consistency.} Let us cosider the Clausius-Duhem inequality which states that the internal dissipation is non-negative.
In the isothermal case it assumes the reduced form
\begin{equation}\label{cldGeneral}
\delta_{\text{i}} := \frac{1}{2 \ \rho_{\scriptscriptstyle \text{R}}} \tilde{\mathbf T} : \dot{\mathbf C} - \dot{\psi} \geq 0.
\end{equation}
Taking into account that $\psi_{\text{el}}$ and $\psi_{\text{kin}}$ are isotropic functions,
after some algebraic computations (cf., for example, \cite{ShutovKreissig2008}) we arrive at
the following specified form of the Clausius-Duhem inequality
\begin{equation}\label{cldLong}
\rho_{\scriptscriptstyle \text{R}} \delta_{\text{i}} =  (\hat{\mathbf{C}}_{\text{e}} \hat{\mathbf S} - \hat{\mathbf X}) : {\hat{\mathbf{D}}}_{\text{cr}}
 + \ (\check{\mathbf{C}}_{\text{ie}} \check{\mathbf X}) : {\check{\mathbf{D}}}_{\text{ii}}
 - \frac{\partial \psi(\hat{\mathbf{C}}_{\text{e}}, \check{\mathbf{C}}_{\text{ie}}, \omega)}{\partial \omega} \ \dot{\omega} \geq 0.
\end{equation}
For the presentation it is convenient to introduce the following
abbreviations
\begin{equation}\label{drforc}
\hat{\mathbf \Sigma} := \hat{\mathbf{C}}_{\text{e}} \hat{\mathbf S} -
\hat{\mathbf X}, \quad
\check{\mathbf \Xi} := \check{\mathbf{C}}_{\text{ie}} \check{\mathbf X}, \quad
Y := \frac{\partial \psi(\hat{\mathbf{C}}_{\text{e}}, \check{\mathbf{C}}_{\text{ie}}, \omega)}{\partial \omega}.
\end{equation}
Here, $\hat{\mathbf \Sigma}$ represents the effective stress operating on $\hat{\mathcal{K}}$,
$\check{\mathbf \Xi}$ is the Mandel-like backstress, operating on $\check{\mathcal{K}}$, and
$Y \leq 0$ is a scalar energy release rate.
Substituting these into \eqref{cldLong} we obtain the Clausius-Duhem inequality in a compact form
\begin{equation}\label{cldShort}
\rho_{\scriptscriptstyle \text{R}} \delta_{\text{i}} =  \hat{\mathbf \Sigma} : {\hat{\mathbf{D}}}_{\text{cr}}
 + \ \check{\mathbf \Xi} : {\check{\mathbf{D}}}_{\text{ii}}
 - Y \ \dot{\omega} \geq 0.
\end{equation}
Now we postulate the evolution equations governing the flows ${\hat{\mathbf{D}}}_{\text{cr}}$, ${\check{\mathbf{D}}}_{\text{ii}}$, and
$\dot{\omega}$ in such a way as to guarantee the inequality \eqref{cldShort}.
Even more, we will show that $\hat{\mathbf \Sigma} : {\hat{\mathbf{D}}}_{\text{cr}} \geq 0$,
$\check{\mathbf \Xi} : {\check{\mathbf{D}}}_{\text{ii}} \geq 0$, and $Y \ \dot{\omega} \leq 0$.
Anologously to \eqref{equivStresslamb}, \eqref{equivStressDire}, and \eqref{CreepEvolut} we postulate
\begin{equation}\label{CreepEvolutLage1}
\sigma^{\lambda}_{\text{eq}} := \alpha^{\lambda}_1 \sigma_{\text{reg max}} (\hat{\mathbf \Sigma}) +
\alpha^{\lambda}_2 \ \sqrt{\frac{3}{2}} \ \| \hat{\mathbf \Sigma}^{\text{D}}  \| +
(1 - \alpha^{\lambda}_1 - \alpha^{\lambda}_2) \ \text{tr}(\hat{\mathbf \Sigma}),
\end{equation}
\begin{equation}\label{CreepEvolutLage2}
\sigma_{\text{eq}} := \alpha \ \frac{3}{2} \ \sigma_{\text{reg max}}(\hat{\mathbf \Sigma}^{\text{D}})  +
(1-\alpha) \ \sqrt{\frac{3}{2}} \ \| \hat{\mathbf \Sigma}^{\text{D}} \|,
\end{equation}
\begin{equation}\label{CreepEvolutLage3}
{\hat{\mathbf{D}}}_{\text{cr}} = \lambda (\sigma^{\lambda}_{\text{eq}},\omega)
\frac{\partial \sigma_{\text{eq}}}{\partial \hat{\mathbf \Sigma}},
\end{equation}
where the weighting coefficients $\alpha^{\lambda}_1$, $\alpha^{\lambda}_2$, and $\alpha$
play the same role as in the small strain case; the function $\lambda (\sigma^{\lambda}_{\text{eq}},\omega)$
is defined by \eqref{KachanRabot} in combination with one of
the equations \eqref{CreepRateNorton}--\eqref{CreepRateGarofalo}.
Since $\sigma_{\text{eq}}$ is a convex function of $\hat{\mathbf \Sigma}$ and $\lambda (\sigma^{\lambda}_{\text{eq}},\omega) \geq 0$,
\eqref{CreepEvolutLage3} immediately yields $\hat{\mathbf \Sigma} : {\hat{\mathbf{D}}}_{\text{cr}} \geq 0$.
Next, analogously to \eqref{KinematHardRecov} we postulate the following flow rule on the substructural level
\begin{equation}\label{KinematHardRecovLS}
{\check{\mathbf{D}}}_{\text{ii}} = \varkappa_{\text{dynam}} \
 \|  {\hat{\mathbf{D}}}_{\text{cr}}  \|  \ \check{\mathbf \Xi}^{\text{D}} + \varkappa_{\text{stat}} \ \check{\mathbf \Xi}^{\text{D}},
\end{equation}
where $\varkappa_{\text{dynam}}$ and $\varkappa_{\text{stat}}$
are the material parameters governing the dynamic and static recovery, respectively.
Since these parameters are non negative, we have $\check{\mathbf \Xi} : {\check{\mathbf{D}}}_{\text{ii}} \geq 0$.
Finally, following \eqref{equivStressOmeg} we introduce the damage-controlling equivalent stress
\begin{equation}\label{equivStressOmegLar}
\sigma^{\omega}_{\text{eq}} := \alpha^{\omega}_1 \sigma_{\text{reg max}} (\hat{\mathbf \Sigma}) +
\alpha^{\omega}_2 \ \sqrt{\frac{3}{2}} \  \| \hat{\mathbf \Sigma}^{\text{D}}  \| +
(1 - \alpha^{\omega}_1 - \alpha^{\omega}_2) \ \text{tr}(\hat{\mathbf \Sigma}).
\end{equation}
The corresponding damage evolution is then given by the scalar equation \eqref{equivStressOmegEv}
which reads $\dot{\omega} = B (1-\omega)^{-l} (\sigma^{\omega}_{\text{eq}})^k$.
For $B \geq 0$ we have $\dot{\omega} \geq 0$. Thus, since $Y \leq 0$, we arrive at $Y \ \dot{\omega} \leq 0$.
Therefore, the proposed finite-strain creep model is thermodynamically consistent.

Since $\sigma_{\text{eq}}$ depends on $\hat{\mathbf \Sigma}^{\text{D}}$, the flow rule \eqref{CreepEvolutLage3}
yields incompressible flow: $\text{tr}{\hat{\mathbf{D}}}_{\text{cr}} = 0$.
Obviously, the substructural flow rule \eqref{KinematHardRecovLS} yields an incompressible flow as well:
$\text{tr}{\check{\mathbf{D}}}_{\text{ii}} = 0$.
Note that the finite-strain version of the creep model contains
the same number of material parameters as its small-strain counterpart presented in the previous subsection.

\textbf{Transformation to the reference configuration.} The flow rules \eqref{CreepEvolutLage3} and \eqref{KinematHardRecovLS} are formulated
on fictitious configurations $\hat{\mathcal{K}}$ and $\check{\mathcal{K}}$.
Let us transform these equations to the reference configuration $\widetilde{\mathcal{K}}$.
First, we note that the similarity transformation $(\cdot) \mapsto
\mathbf F^{\text{T}}_{\text{cr}} (\cdot) \mathbf F^{-\text{T}}_{\text{cr}}$
 maps the effective stress $\hat{\mathbf \Sigma}$
to its non-symmetric counterpart $\tilde{\mathbf \Sigma}$ operating on the reference configuration:
\begin{equation}\label{FiSimilarity}
\tilde{\mathbf \Sigma} = \mathbf F^{\text{T}}_{\text{cr}} \ \hat{\mathbf \Sigma} \ \mathbf F^{-\text{T}}_{\text{cr}}, \quad \text{where} \quad
\tilde{\mathbf \Sigma} := \mathbf C \tilde{\mathbf T} - \mathbf C_{\text{cr}} \tilde{\mathbf X} \notin Sym.
\end{equation}
Since the similarity preserves the invariants, we have
\begin{equation}\label{SimilarityInvariants}
\text{eigenvalues of} \ \hat{\mathbf \Sigma} = \text{eigenvalues of} \ \tilde{\mathbf \Sigma}, \quad
\sigma_{\text{reg max}}(\hat{\mathbf \Sigma}) = \sigma_{\text{reg max}}(\tilde{\mathbf \Sigma}), \quad
\text{tr} \hat{\mathbf \Sigma} = \text{tr} \tilde{\mathbf \Sigma}.
\end{equation}
The Frobenius norm of $\hat{\mathbf \Sigma}^{\text{D}}$ is represented now as follows (cf. \cite{Kiesling})
\begin{equation}\label{NormOfDevPart}
\| \hat{\mathbf \Sigma}^{\text{D}}  \|  = \mathfrak{N} ( \hat{\mathbf \Sigma}^{\text{D}} ) = \mathfrak{N} ( \tilde{\mathbf \Sigma}^{\text{D}} ), \quad
\text{where} \quad \mathfrak{N} (\mathbf{A}):= \sqrt{ \text{tr}(\mathbf{A} \mathbf{A} )} \ \ \text{for all} \ \ \mathbf{A}.
\end{equation}
Note that the introduced function $\mathfrak{N} (\cdot)$ is not a norm. Nevertheless, $\mathfrak{N} ( \tilde{\mathbf \Sigma}^{\text{D}} )$
is still a physically reasonable quantity since it is equal to the norm of the driving force $\| \hat{\mathbf \Sigma}^{\text{D}}  \|$.
Thus, the previously introduced equivalent stresses take the following form on the reference configuration
\begin{equation}\label{EqStresReferenceLamb}
\sigma^{\lambda}_{\text{eq}} = \alpha^{\lambda}_1 \sigma_{\text{reg max}} (\tilde{\mathbf \Sigma}) +
\alpha^{\lambda}_2 \ \sqrt{\frac{3}{2}} \  \mathfrak{N} ( \tilde{\mathbf \Sigma}^{\text{D}} ) +
(1 - \alpha^{\lambda}_1 - \alpha^{\lambda}_2) \ \text{tr}(\tilde{\mathbf \Sigma}),
\end{equation}
\begin{equation}\label{EqStresReference}
\sigma_{\text{eq}} = \alpha \ \frac{3}{2} \ \sigma_{\text{reg max}}(\tilde{\mathbf \Sigma}^{\text{D}})  +
(1-\alpha) \ \sqrt{\frac{3}{2}} \ \mathfrak{N} (\tilde{\mathbf \Sigma}^{\text{D}}),
\end{equation}
\begin{equation}\label{EqStresReferenceOmega}
\sigma^{\omega}_{\text{eq}} = \alpha^{\omega}_1 \sigma_{\text{reg max}} (\tilde{\mathbf \Sigma}) +
\alpha^{\omega}_2 \ \sqrt{\frac{3}{2}} \  \mathfrak{N} (\tilde{\mathbf \Sigma}^{\text{D}} ) +
(1 - \alpha^{\omega}_1 - \alpha^{\omega}_2) \ \text{tr}(\tilde{\mathbf \Sigma}).
\end{equation}
In the same way we consider the similarity $(\cdot) \mapsto
\mathbf F^{\text{T}}_{\text{ii}} (\cdot) \mathbf F^{-\text{T}}_{\text{ii}}$. This
transformation maps the Mandel-like backstress $\check{\mathbf \Xi}$ to its referential counterpart $\tilde{\mathbf \Xi}$:
\begin{equation}\label{ReferenBackstress}
\tilde{\mathbf \Xi} := \mathbf C_{\text{cr}} \tilde{\mathbf X} \notin Sym, \quad
\tilde{\mathbf \Xi} = \mathbf F^{\text{T}}_{\text{ii}} \  \check{\mathbf \Xi} \ \mathbf F^{-\text{T}}_{\text{ii}}.
\end{equation}
In order to transform the evolution equation \eqref{CreepEvolutLage3}, we note that
\begin{equation}\label{HelpTransform}
\dot{\mathbf{C}}_{\text{cr}} = 2 \ \mathbf F^{\text{T}}_{\text{cr}} \ {\hat{\mathbf{D}}}_{\text{cr}} \ \mathbf F_{\text{cr}},
\quad \mathbf F^{-1}_{\text{cr}} \ \frac{\partial \sigma_{\text{eq}}}{\partial \hat{\mathbf \Sigma}} \ \mathbf F_{\text{cr}}
 = \frac{\partial \sigma_{\text{eq}}}{\partial \tilde{\mathbf \Sigma}}.
\end{equation}
Applying the creep-induced pull-back $ (\cdot) \mapsto \mathbf F^{\text{T}}_{\text{cr}} (\cdot) \mathbf F_{\text{cr}}$
to both sides of \eqref{CreepEvolutLage3} and taking \eqref{HelpTransform} into account we arrive at
\begin{equation}\label{ResultCcr}
\dot{\mathbf{C}}_{\text{cr}} = 2 \ \lambda (\sigma^{\lambda}_{\text{eq}},\omega) \ \mathbf{C}_{\text{cr}} \
\frac{\partial \sigma_{\text{eq}}}{\partial \tilde{\mathbf \Sigma}} =
2 \ \lambda (\sigma^{\lambda}_{\text{eq}},\omega)
\Big( \frac{\partial \sigma_{\text{eq}}}{\partial \tilde{\mathbf \Sigma}} \Big)^{\text{T}}  \ \mathbf{C}_{\text{cr}}.
\end{equation}

In particular, for $\alpha =0$ we have
$\sigma_{\text{eq}} =  \sqrt{\frac{3}{2}} \ \| \hat{\mathbf \Sigma}^{\text{D}} \|
 = \sqrt{\frac{3}{2}} \mathfrak{N} (\tilde{\mathbf \Sigma}^{\text{D}}) $ and
thus we restore the finite-strain version of the $J_2$ flow rule (cf. \cite{ShutovKreissig2008, Kiesling}):
\begin{equation}\label{J2FlowCcr}
\frac{\partial \mathfrak{N} (\tilde{\mathbf \Sigma}^{\text{D}})}{\partial \tilde{\mathbf \Sigma}} =
(\tilde{\mathbf \Sigma}^{\text{D}})^{\text{T}}
 \quad \Rightarrow \quad
\dot{\mathbf{C}}_{\text{cr}} = \sqrt{6}
\frac{ \lambda}{\mathfrak{N} (\tilde{\mathbf \Sigma}^{\text{D}})} \
 \tilde{\mathbf \Sigma}^{\text{D}} \ \mathbf{C}_{\text{cr}} \quad \text{for} \quad \alpha =0.
\end{equation}

Further, to transform equation \eqref{KinematHardRecovLS} we note that the
norm of the creep strain rate $\|  {\hat{\mathbf{D}}}_{\text{cr}}  \|$ can be written in
Lagrangian description as follows
\begin{equation}\label{NormCreepRate}
\|  {\hat{\mathbf{D}}}_{\text{cr}}  \| = \frac{1}{2} \mathfrak{N} ( \mathbf{C}^{-1}_{\text{cr}} \ \dot{\mathbf{C}}_{\text{cr}} ) =
\frac{1}{2} \| \mathbf{C}^{-1/2}_{\text{cr}} \ \dot{\mathbf{C}}_{\text{cr}} \ \mathbf{C}^{-1/2}_{\text{cr}}  \|.
\end{equation}
Moreover, using the identity $\text{tr} \tilde{\mathbf \Xi} = \text{tr} \check{\mathbf \Xi}$, after some algebraic
computations we arrive at (cf. \cite{ShutovKreissig2008})
\begin{equation}\label{HelpTransform2}
\dot{\mathbf{C}}_{\text{ii}} = 2 \ \mathbf F^{\text{T}}_{\text{ii}} \ {\check{\mathbf{D}}}_{\text{ii}} \ \mathbf F_{\text{ii}}, \quad
\quad \mathbf F^{\text{T}}_{\text{ii}} \ \check{\mathbf \Xi}^{\text{D}} \ \mathbf F_{\text{ii}} = \tilde{\mathbf \Xi}^{\text{D}} \ \mathbf C_{\text{ii}}.
\end{equation}
Combining \eqref{NormCreepRate}, \eqref{HelpTransform2} and \eqref{KinematHardRecovLS}
we arrive at equation which governs the backstress saturation
\begin{equation}\label{KinematHardRefConf}
\dot{\mathbf{C}}_{\text{ii}} = \big( \varkappa_{\text{dynam}} \
 \mathfrak{N} ( \mathbf{C}^{-1}_{\text{cr}} \ \dot{\mathbf{C}}_{\text{cr}} )
 + 2 \varkappa_{\text{stat}} \big) \ \tilde{\mathbf \Xi}^{\text{D}} \  \mathbf{C}_{\text{ii}}.
\end{equation}
The flow rules \eqref{ResultCcr} and \eqref{KinematHardRefConf} are incompressible; under appropriate
initial conditions $\mathbf{C}_{\text{cr}}|_{t=0} = \mathbf{C}^0_{\text{cr}}$ and
$\mathbf{C}_{\text{ii}}|_{t=0} = \mathbf{C}^0_{\text{ii}}$ we have:
$\det \mathbf{C}_{\text{cr}} = \det \mathbf{C}_{\text{ii}} = 1$.
Since the free energy functions $\psi_{\text{el}}(\hat{\mathbf{C}}_{\text{e}}, \omega)$ and
$\psi_{\text{kin}}(\check{\mathbf{C}}_{\text{ie}}, \omega)$ are isotropic, they can be rewritten in the following way
\begin{equation}\label{FreeEnergTrafo}
\psi_{\text{el}}(\hat{\mathbf{C}}_{\text{e}}, \omega) =
\psi_{\text{el}}(\mathbf C \mathbf C^{-1}_{\text{cr}}, \omega), \quad
\psi_{\text{kin}}(\check{\mathbf{C}}_{\text{ie}}, \omega) =
\psi_{\text{kin}}(\mathbf C_{\text{cr}} \ \mathbf C^{-1}_{\text{ii}}, \omega).
\end{equation}
Finally, using this result, the hyperelastic relations \eqref{Hyperelas}
are transformed to the reference configuration; now they provide the
second Piola-Kirchhoff stress $\tilde{\mathbf T}$
and the backstress $\tilde{\mathbf X}$ (cf. eq. (48) in Ref. \cite{ShutovKreissig2008})
\begin{equation}\label{HyperelasRef}
\tilde{\mathbf T}  =
2 \rho_{\scriptscriptstyle \text{R}}
\frac{\displaystyle \partial \psi_{\text{el}}(\mathbf C \mathbf C^{-1}_{\text{cr}}, \omega)}
{\displaystyle \partial \mathbf{C}}\big|_{\mathbf C_{\text{cr}} = \text{const}}, \quad
\tilde{\mathbf X}  =
2 \rho_{\scriptscriptstyle \text{R}}
\frac{\displaystyle \partial \psi_{\text{kin}}(\mathbf C_{\text{cr}} \ \mathbf C^{-1}_{\text{ii}}, \omega)}
{\displaystyle \partial \mathbf C_{\text{cr}}}\big|_{\mathbf C_{\text{ii}} = \text{const}}.
\end{equation}
In particular, if the neo-Hookean potentials \eqref{neoHooke1} and \eqref{neoHooke2} are employed, we have
\begin{equation}\label{StrBackHooke}
\tilde{\mathbf T} = (1-\omega) \Big[ \frac{\displaystyle k}{\displaystyle 10} \
\big( (\text{det} \mathbf C)^{5/2}-(\text{det} \mathbf C)^{-5/2} \big) \
\mathbf C^{-1} + \mu \ \mathbf C^{-1} (\overline{\mathbf C}
\mathbf C_{\text{cr}}^{-1})^{\text{D}}\Big], \quad
  \tilde{\mathbf X} = (1-\omega) \frac{c}{2} \ \mathbf C_{\text{cr}}^{-1} (\mathbf C_{\text{cr}} \mathbf C_{\text{ii}}^{-1})^{\text{D}}.
\end{equation}

The system of constitutive equations is summarized in Table \ref{table1}.
As already shown, the model is thermodynamically admissible.
The objectivity of the model follows from the fact that the second Piola-Kirchoff stress
depends solely on the history of the right Cauchy-Green tensor (and some initial conditions). Moreover,
following the procedure presented in \cite{ShutovPfeiffer}, an important property of the model can be proved:
upon the isochoric change of the reference configuration the model predicts the same Cauchy stresses if the
initial conditions imposed on $\mathbf{C}_{\text{cr}}$ and $\mathbf{C}_{\text{ii}}$ are properly transformed. This property is referred
to as a weak invariance \cite{ShutovAnalysisOfSome}.

\begin{table}
\caption{Summary of the cyclic creep model formulated on the reference configuration}
\begin{tabular}{|l l|}
\hline
$\tilde{\mathbf T}  =
2 \rho_{\scriptscriptstyle \text{R}}
\frac{\displaystyle \partial \psi_{\text{el}}(\mathbf C \mathbf C^{-1}_{\text{cr}}, \omega)}
{\displaystyle \partial \mathbf{C}}\big|_{\mathbf C_{\text{cr}} = \text{const}} $, &
$\tilde{\mathbf X}  =
2 \rho_{\scriptscriptstyle \text{R}}
\frac{\displaystyle \partial \psi_{\text{kin}}(\mathbf C_{\text{cr}} \ \mathbf C^{-1}_{\text{ii}}, \omega)}
{\displaystyle \partial \mathbf C_{\text{cr}}}\big|_{\mathbf C_{\text{ii}} = \text{const}} $, \\
$\tilde{\mathbf \Sigma} := \mathbf C \tilde{\mathbf T} - \mathbf C_{\text{cr}} \tilde{\mathbf X}$, \ \
$\tilde{\mathbf \Xi} := \mathbf C_{\text{cr}} \tilde{\mathbf X} $, &
$\mathfrak{N} (\mathbf{A}):= \sqrt{ \text{tr}(\mathbf{A}^2 )}$, \\
$\sigma^{\lambda}_{\text{eq}} = \alpha^{\lambda}_1 \sigma_{\text{reg max}} (\tilde{\mathbf \Sigma}) +
\alpha^{\lambda}_2 \ \sqrt{\frac{3}{2}} \  \mathfrak{N} ( \tilde{\mathbf \Sigma}^{\text{D}} ) +
(1 - \alpha^{\lambda}_1 - \alpha^{\lambda}_2) \ \text{tr}(\tilde{\mathbf \Sigma})$, &
$\sigma_{\text{eq}} = \alpha \ \frac{3}{2} \ \sigma_{\text{reg max}}(\tilde{\mathbf \Sigma}^{\text{D}})  +
(1-\alpha) \ \sqrt{\frac{3}{2}} \ \mathfrak{N} (\tilde{\mathbf \Sigma}^{\text{D}} )$, \\
$\sigma^{\omega}_{\text{eq}} = \alpha^{\omega}_1 \sigma_{\text{reg max}} (\tilde{\mathbf \Sigma}) +
\alpha^{\omega}_2 \ \sqrt{\frac{3}{2}} \  \mathfrak{N} (\tilde{\mathbf \Sigma}^{\text{D}} ) +
(1 - \alpha^{\omega}_1 - \alpha^{\omega}_2) \ \text{tr}(\tilde{\mathbf \Sigma})$, & \\
$\dot{\mathbf{C}}_{\text{cr}} =
2 \ \lambda (\sigma^{\lambda}_{\text{eq}},\omega)
\Big( \frac{\displaystyle \partial \sigma_{\text{eq}}}{\displaystyle \partial \tilde{\mathbf \Sigma}} \Big)^{\text{T}}  \ \mathbf{C}_{\text{cr}}$,
 & $\mathbf C_{\text{cr}}|_{t=0} = \mathbf C_{\text{cr}}^0$, $\det \mathbf C_{\text{cr}}^0 =1$,  \\
 $\dot{\mathbf{C}}_{\text{ii}} = \big( \varkappa_{\text{dynam}} \
 \mathfrak{N} ( \mathbf{C}^{-1}_{\text{cr}} \ \dot{\mathbf{C}}_{\text{cr}} )
 + 2 \varkappa_{\text{stat}} \big) \ \tilde{\mathbf \Xi}^{\text{D}} \  \mathbf{C}_{\text{ii}}$, &
$\mathbf C_{\text{ii}}|_{t=0} = \mathbf C_{\text{ii}}^0$, $\det \mathbf C_{\text{ii}}^0 =1$,  \\
$\dot{\omega} = B (1-\omega)^{-l} (\sigma^{\omega}_{\text{eq}})^k$, & $\omega|_{t=0} = \omega^0 $ \\
 \hline
\end{tabular}
\label{table1}
\end{table}

\section{Numerical implementation}

We note that the exact solution to the evolution equations
\eqref{ResultCcr} and \eqref{KinematHardRefConf} exhibits the following geometric property
\begin{equation}\label{geopro}
\mathbf C_{\text{cr}}, \mathbf C_{\text{ii}} \in \mathbb{M}, \quad \text{where} \quad
\mathbb{M} := \big\{ \mathbf B \in Sym: \text{det} \mathbf B =1 \big\}.
\end{equation}
Therefore, we say that we are dealing with a system of ordinary differential equations on the manifold $\mathbb{M} \times \mathbb{M}$.
Obviously, the symmetry condition $\mathbf C_{\text{cr}}, \mathbf C_{\text{ii}} \in Sym$ should be satisfied
by any numerical scheme, since $\mathbf C_{\text{cr}}$ and $\mathbf C_{\text{ii}}$
represent some metric tensors of Cauchy-Green type (see \eqref{RCG}). Next, the exact preservation of the inelastic
incompressibility $\det(\mathbf C_{\text{cr}})=\det(\mathbf C_{\text{ii}})=1$
is needed to suppress the accumulation of the numerical error (see the discussion in \cite{ShutovKreissigGeometr}).
In this section we propose a numerical procedure which will exactly satisfy the geometric properties \eqref{geopro}.
In the current study, the numerical implementation of the model is carried out for the
neo-Hookean potentials \eqref{neoHooke1} and \eqref{neoHooke2}.

Let us consider a typical time step $t_n \mapsto t_{n+1}$. The current time step size
 is denoted by $\Delta t := t_{n+1} - t_{n}>0$.
We assume that the current value of the right Cauchy-Green tensor ${}^{n+1} \mathbf C$
is known. Moreover, the values of the internal variables are given at the previous time step by
${}^n \mathbf C_{\text{cr}}$, ${}^n \mathbf C_{\text{ii}}$, and ${}^n \omega$. In order to compute
the actual stress tensor ${}^{n+1}\tilde{\mathbf T}$ we update the internal variables by integrating
the corresponding evolution equations.
The evolution equations
\eqref{ResultCcr} and \eqref{KinematHardRefConf}
which govern the inelastic flow are discretized using an implicit scheme, damage evolution \eqref{equivStressOmegEv}
is treated by the explicit Euler method.
First, we rewrite \eqref{ResultCcr} in a more compact form
\begin{equation}\label{CompactEvolution}
\dot{\mathbf{C}}_{\text{cr}} = \mathbf{f}_{\text{cr}}(\mathbf C, \mathbf C_{\text{cr}},
\mathbf C_{\text{ii}}, \omega)
  \ \mathbf{C}_{\text{cr}}, \quad \text{where} \quad
\mathbf{f}_{\text{cr}}(\mathbf C, \mathbf C_{\text{cr}},
\mathbf C_{\text{ii}}, \omega) := 2 \ \lambda (\sigma^{\lambda}_{\text{eq}},\omega)
\Big( \frac{\partial \sigma_{\text{eq}}}{\partial \tilde{\mathbf \Sigma}} \Big)^{\text{T}}.
\end{equation}
Unfortunately, due to its linear structure, the Euler-Backward method (EBM) violates
the incompressibility condition. For that reason the following modified version of the EBM is considered
here for the implicit discretization of \eqref{CompactEvolution} (cf. eq. (74) in Ref. \cite{ShutovKreissig2008}).
\begin{equation}\label{ModifEBMCcr}
{}^{n+1} \mathbf C_{\text{cr}} = \overline{ \big[ \mathbf{1} - \Delta t \
\mathbf{f}_{\text{cr}}({}^{n+1} \mathbf C, {}^{n+1} \mathbf C_{\text{cr}},
{}^{n+1} \mathbf C_{\text{ii}}, {}^{n} \omega) \big]^{-1}}  \ {}^{n} \mathbf C_{\text{cr}}.
\end{equation}
It can be shown that the classical Euler-Backward discretization and its modification \eqref{ModifEBMCcr}
automatically preserve the symmetry condition $\mathbf C_{\text{cr}} \in Sym$ (see Appendix A).
Thus, the following symmetrized modification is equivalent to \eqref{ModifEBMCcr} (cf. eq. (75) in Ref. \cite{ShutovKreissig2008})
\begin{equation}\label{ModifEBMCcrSym}
{}^{n+1} \mathbf C_{\text{cr}} = \overline{ \text{sym}  \Big\{ \big[ \mathbf{1} - \Delta t \
\mathbf{f}_{\text{cr}}({}^{n+1} \mathbf C, {}^{n+1} \mathbf C_{\text{cr}},
{}^{n+1} \mathbf C_{\text{ii}}, {}^{n} \omega) \big]^{-1}  \ {}^{n} \mathbf C_{\text{cr}} \Big\} }.
\end{equation}
Obviously, this scheme exactly preserves the geometric
property: ${}^{n+1} \mathbf{C}_{\text{cr}} \in \mathbb{M}$.
Further, we recall that a neo-Hookean potential \eqref{neoHooke2} is adopted for $\psi_{\text{kin}}$ in the current study.
Therefore, $\eqref{StrBackHooke}_2$ is valid and the evolution equation
\eqref{KinematHardRefConf} takes the following specific form
\begin{equation}\label{KinematHardSpecific}
\dot{\mathbf{C}}_{\text{ii}} = (1-\omega) \ c \ \big( \frac{1}{2} \varkappa_{\text{dynam}} \
 \mathfrak{N} ( \mathbf{C}^{-1}_{\text{cr}} \ \dot{\mathbf{C}}_{\text{cr}} )
 + \varkappa_{\text{stat}} \big) \ (\mathbf C_{\text{cr}} \mathbf C_{\text{ii}}^{-1})^{\text{D}} \  \mathbf{C}_{\text{ii}}.
\end{equation}
Assume that the creep strain rate is approximated by a constant within the time step:
\begin{equation}\label{CreepRateApprox}
\mathfrak{N} ( \mathbf{C}^{-1}_{\text{cr}} \ \dot{\mathbf{C}}_{\text{cr}} )
 \approx \frac{1}{\Delta t}
\mathfrak{N} \big({}^{n+1} \mathbf{C}^{-1}_{\text{cr}}   ( {}^{n+1} \mathbf{C}_{\text{cr}}  - {}^{n} \mathbf{C}_{\text{cr}}  ) \big).
\end{equation}
Substituting this into the right-hand side of \eqref{KinematHardSpecific} we obtain an
evolution equation, which has exactly the same structure as for the
multiplicative finite-strain Maxwell fluid of Simo and Miehe (cf. eq. (14) in Ref. \cite{ShutovLandgraf}).
For this version of the Maxwell fluid an explicit update formula is available
(cf. eq. (29) in Ref. \cite{ShutovLandgraf}). In current notations, this update
formula reads
\begin{equation}\label{EcplicitUpdate}
{}^{n+1} \mathbf{C}_{\text{ii}} = \overline{{}^{n} \mathbf{C}_{\text{ii}} +  \ (1- {}^{n} \omega) \ c \ \Big(
\frac{1}{2} \varkappa_{\text{dynam}} \
\mathfrak{N} \big({}^{n+1} \mathbf{C}^{-1}_{\text{cr}}
( {}^{n+1} \mathbf{C}_{\text{cr}}  - {}^{n} \mathbf{C}_{\text{cr}}  ) \big)
 + \Delta t \ \varkappa_{\text{stat}}\Big) \ {}^{n+1} \mathbf{C}_{\text{cr}}}.
\end{equation}
Obviously, this formula is a geometric integrator: ${}^{n+1} \mathbf{C}_{\text{ii}} \in \mathbb{M}$;
it yields ${}^{n+1} \mathbf{C}_{\text{ii}}$ as an \emph{explicit}
function: ${}^{n+1} \mathbf{C}_{\text{ii}} = \mathfrak{C}_{\text{ii}} ({}^{n+1} \mathbf{C}_{\text{cr}})$.
Substituting explicit update \eqref{EcplicitUpdate}
into \eqref{ModifEBMCcrSym}, the overall procedure boils down to the solution of the following
nonlinear equation with respect to unknown tensor ${}^{n+1} \mathbf C_{\text{cr}}$
\begin{equation}\label{IntegFinal}
{}^{n+1} \mathbf C_{\text{cr}} = \overline{ \text{sym}  \Big\{ \big[ \mathbf{1} - \Delta t \
\mathbf{f}_{\text{cr}}({}^{n+1} \mathbf C, {}^{n+1} \mathbf C_{\text{cr}},
\mathfrak{C}_{\text{ii}} ({}^{n+1} \mathbf{C}_{\text{cr}}), {}^{n} \omega) \big]^{-1}  \ {}^{n} \mathbf C_{\text{cr}} \Big\} }.
\end{equation}
This equation is solved at each time step iteratively by the Newton-Raphson method.

\textbf{Remark 4.}
In the special case of the flow rule \eqref{J2FlowCcr} which corresponds to $\alpha =0$,
the evolution equation governing $\mathbf C_{\text{cr}}$ has the same
structure as for the model of multiplicative
viscoplasticity proposed in \cite{ShutovKreissig2008}.
An explicit update formula is described in \cite{Shutov2016} for this evolution equation; this explicit solution
can be used for the presented creep model as well. As a result, the overall time stepping can be reduced
to the solution of a single scalar equation (cf. \cite{Shutov2016}). $\Box$

After ${}^{n+1} \mathbf C_{\text{cr}}$ and ${}^{n+1} \mathbf C_{\text{ii}} =
\mathfrak{C}_{\text{ii}} ({}^{n+1} \mathbf{C}_{\text{cr}})$
are found, the damage evolution
\eqref{equivStressOmegEv} is integrated using
the explicit Euler method
\begin{equation}\label{DamageEvolut}
{}^{n+1} \omega = {}^{n} \omega + \Delta t \ B (1- {}^{n} \omega)^{-l} ({}^{n} \sigma^{\omega}_{\text{eq}})^k.
\end{equation}

\section{Validation of the model}

\begin{table}[h]
\caption{Material parameters.}
\begin{tabular}{| l | l | l|l|}
\hline
parameter & value       &  brief explanation   &   equation      \\ \hline \hline
$k$            & 73500  MPa     &  bulk modulus of intact material &   \eqref{neoHooke1}    \\ \hline
$\mu$        & 28200   MPa    &   shear modulus of intact material &  \eqref{neoHooke1}     \\ \hline
$c$          & 7550   MPa    &   shear modulus of intact substructure &  \eqref{neoHooke2}     \\ \hline
$A$          & $1.185 \cdot 10^{-13} \ \ \text{h}^{-1}   $    &  parameter of Norton's law &  \eqref{CreepRateNorton}     \\ \hline
$n$          & 5  [-]                        &  parameter of Norton's law (exponent)&  \eqref{CreepRateNorton}     \\ \hline
$m$          & 30 [-]                        &  impact of damage &  \eqref{KachanRabot}     \\ \hline
$\varkappa_{\text{dyam}}$    & 0.055  $\text{MPa}^{-1}$    &   dynamic recovery coefficient &  \eqref{KinematHardRecovLS}  \\ \hline
$\varkappa_{\text{stat}}$    & 0.0  $\text{MPa}^{-1} \text{h}^{-1}$    &   static recovery coefficient &  \eqref{KinematHardRecovLS}  \\ \hline
$\alpha$    & 0.0  [-]    &   weighting coefficient for $\sigma_{\text{eq}}$ &  \eqref{EqStresReference}  \\ \hline
$\alpha^{\lambda}_1$    & 0.0  [-]    &   weighting coefficient for $\sigma^{\lambda}_{\text{eq}} $ &  \eqref{EqStresReferenceLamb}  \\ \hline
$\alpha^{\lambda}_2$    & 1.0  [-]    &   weighting coefficient for $\sigma^{\lambda}_{\text{eq}} $ &  \eqref{EqStresReferenceLamb}  \\ \hline
$\alpha^{\omega}_1$     & 0.0  [-]    &   weighting coefficient for $\sigma^{\omega}_{\text{eq}} $  &  \eqref{EqStresReferenceOmega}  \\ \hline
$\alpha^{\omega}_2$     & 1.0  [-]    &   weighting coefficient for $\sigma^{\omega}_{\text{eq}} $  &  \eqref{EqStresReferenceOmega}  \\ \hline
$l$     & 0.0  [-]    &  damage evolution parameter  &  \eqref{equivStressOmegEv}  \\ \hline
$k$     & 5.0  [-]    &  damage evolution parameter  &  \eqref{equivStressOmegEv}  \\ \hline
$\omega_0$     & 0.01  [-]    &  initial damage  &  $\eqref{InitCond}_3$  \\ \hline
\end{tabular}
\label{tab2}
\end{table}

The finite-strain cyclic creep model presented in this study is implemented into MSC.MARC as a user-defined material subroutine using
the Hypela2 interface.
For the initial validation of the model we simulate a torsion test
performed on a thick-walled tubular specimen made
of the Russian D16T aluminum alloy. The transient creep response of this alloy is
of big interest since it is widely used in aerospace applications.
It corresponds to AlCuMg2 (see the German DIN 1745); it is also
similar to the 24ST4 alloy.\footnote{Another study concerned with this
alloy is presented in \cite{Klebanov}.}

\begin{figure}\centering
\scalebox{0.95}{\includegraphics{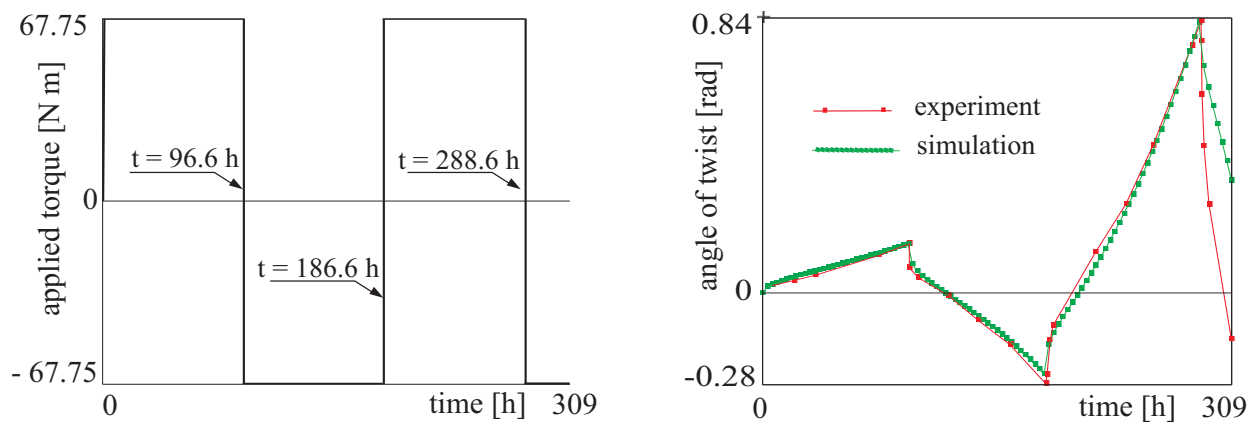}}
\caption{Applied torque as a stepwise function of time (left) and the twist angle as a function of time (right).
The experimental data are taken from \cite{Gorev} and correspond to the constant temperature $T = 250 {}^{\text{o}}C$. \label{fig3}}
\end{figure}

\begin{figure}\centering
\scalebox{0.95}{\includegraphics{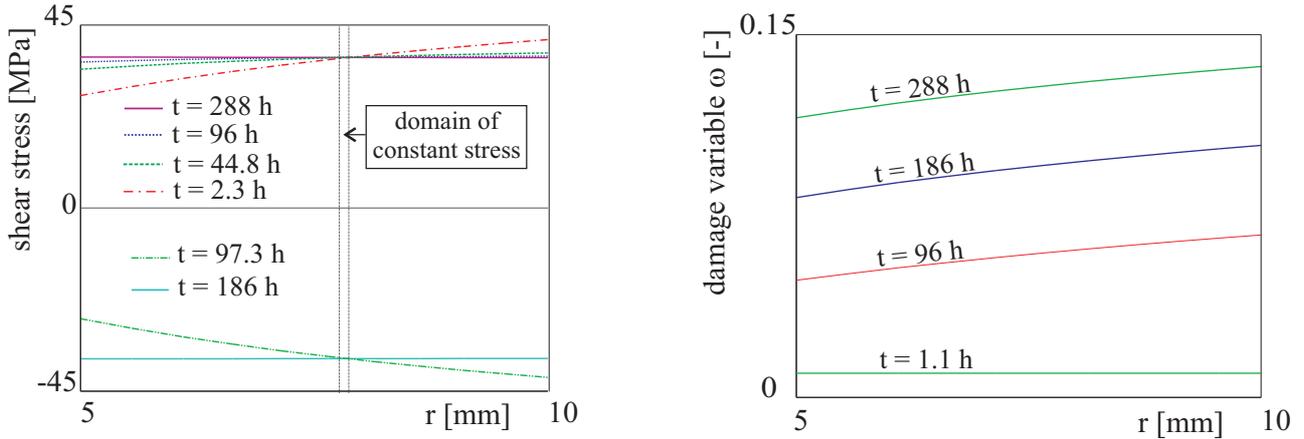}}
\caption{Distribution of the shear stress at different instances of time $t$ over the radius $r$ (left); distribution of the
damage $\omega$ over the radius $r$ at different instances of time $t$ (right). \label{fig4}}
\end{figure}

The dimensions of the thick-walled tubular
specimen in the gage area are as follows: length $L= 70$ mm, inner radius $r_{\text{i}}=5$ mm, outer radius $r_{\text{o}}=10$ mm.
Here we use experimental creep data reported in \cite{Gorev}.
The applied torque is a stepwise constant function of time shown
in Fig. \ref{fig3}(left), the temperature is held constant during the entire process: $T = 250 {}^{\text{o}}C$.
The experimentally measured twist is plotted against time
on Fig. \ref{fig3}(right). The Norton creep law \eqref{CreepRateNorton} is used in this section to simulate the torsion
creep test; the relevant material parameters are summarized in Table \ref{tab2}.\footnote{
The proper parameter identification for the D16T alloy is not the goal of the present study.
In this section we merely demonstrate the ability of the proposed material model
to capture certain mechanical phenomena.}

As can be seen from Fig. \ref{fig3}(right), the transient phases of the non-stationary creep
after load reversals
can be captured by the model with a good accuracy.
The overall effect of the creep damage is apparent; an increased creep strain rate is observed at the final
part of the experiment.
The effect of increasing amplitude of the (instant) elastic strain
is captured by the model as well (cf. Fig. \ref{fig3}(right)). This effect is explained by the deterioration of elastic properties
 in the damaged material. Another important feature described by the model is that the transient stage is getting longer
with progressive damage. Interestingly, this material phenomenon is explained by the current model as an interplay between the
creep damage and the static
  recovery of the backstresses. The computed distribution
  of the shear stress (true stresses are used)
at different instances of time is shown in Fig. \ref{fig4}(left).
As is typical for creep problems, the stress distribution is becoming more uniform with time under constant applied stress.
However, immediately after load reversals the stress distribution becomes inhomogeneous again.

Within the current simulation, the redistribution of stresses occur rather fast: within a few hours the stress distribution becomes
nearly homogeneous (cf. the curve for $t = 2.3$ h shown in Fig. \ref{fig4}(left)).
Therefore, the pronounced transient effect observed in the simulation (cf. \ref{fig3}(right))
is due to the material behaviour, not due to the
heterogeneous character of the stress distribution in the sample.

Examination of Fig. \ref{fig4}(left) shows another interesting feature: Within the
sample there is a domain which is not affected by the redistribution of stresses.
In this domain the absolute value of the shear stress is a constant function of time.
Some authors refer to these stresses as  ``skeletal point stresses" or  ``referential stresses" \cite{Hayhurst1973}.
The skeletal point stresses can be used for a rapid and simple interpretation of the experimental data.

The distribution of the damage  variable $\omega$ is shown in \ref{fig4}(right).
Here, $\omega$ is a smooth and monotonic function of the radius $r$.
Although the stress distribution is nearly homogeneous, the material damage
on the outer side is more pronounced than the damage on the inner side.

\section{Discussion and conclusion}

In the current work we put the main focus on the accurate simulation of
the transient (non-stationary) creep response caused by abrupt
load changes.
A thermodynamically consistent finite-strain model with backstresses is proposed here;
the nonlinear evolution of backstresses
allows both for static and dynamic recovery.
The model is coupled to the classical Kachanov-Rabotnov damage evolution such
that the elastic properties deteriorate with damage.

The current framework allows one to consider the creep strain rate as an isotropic function
of the effective stress. However, in order to reduce the huge manifold of
possible constitutive assumptions, we use the equivalent stress $\sigma_{\text{eq}}$ as a creep potential.
An important feature of the current definition of $\sigma_{\text{eq}}$ is that the maximum positive eigenvalue
is replaced by its regularized (smoothed) counterpart $\sigma_{\text{reg max}}$. This regularization is needed
for numerical reasons to ensure that the creep strain rate is a
continuous function of the applied stress.

The strength-difference effect (also known as a
tension-compression anisotropy) is taken into account by the
evolution equation \eqref{CreepEvolutLage3} since
the corresponding equivalent stress $\sigma^{\lambda}_{\text{eq}}$ is
sensitive to the sign of the uniaxial loading.
Like any other
model with backstresses, for its calibration one needs a series of
experiments with varying stresses, e.g. tests with stress reversal.
The applicability of the creep model is demonstrated using a FEM simulation of the
non-monotonic torsion of a thick-walled tubular sample.
Model predictions are compared to the experimental results obtained
for the D16T aluminum alloy. A good correspondence between experimental
and theoretical results can be achieved using the proposed model.
The evolution of the stress distribution exhibits a ``skeletal point" within the sample.

In the current study, only one backstress tensor is implemented.
The generalization of the model to cover numerous backstresses is obvious (cf. \cite{ShutovIncrem, Shutov2016}).
Further, we recall that the backstress tensor is purely deviatoric in this study.
Generalization to backstresses with nonzero hydrostatic component is straightforward.

The main conclusion of the paper is that the practically important phenomenon of
the non-stationary creep can be described using the nested multiplicative split of the deformation
gradient. Additional multiplicative decompositions can be introduced to capture
the damage-induced porosity \cite{ShutovDuctDamage} and the thermal expansion \cite{Lion, ShutovThermal}
of the material.
Moreover, the advocated here nested multiplicative split seems a reasonable tool for the construction
of a unified model for creep-plasticity interaction.

\begin{acknowledgement}
  The financial support provided by RFBR (grant number 16-08-00713 À and
  15-01-07631) is acknowledged.
\end{acknowledgement}

\section*{Appendix A}

The classical Euler-backward method (EBM) for the evolution equation
\eqref{CompactEvolution} can be written in two equivalent forms:
\begin{equation}\label{ClassicalEBM1}
{}^{n+1} \mathbf C_{\text{cr}} =
{}^{n} \mathbf C_{\text{cr}} + \Delta t \
\mathbf{f}_{\text{cr}}({}^{n+1} \mathbf C, {}^{n+1} \mathbf C_{\text{cr}},
{}^{n+1} \mathbf C_{\text{ii}}, {}^{n} \omega) \ {}^{n+1} \mathbf C_{\text{cr}},
\end{equation}
\begin{equation}\label{ClassicalEBM2}
{}^{n+1} \mathbf C_{\text{cr}} =  \big[ \mathbf{1} - \Delta t \
\mathbf{f}_{\text{cr}}({}^{n+1} \mathbf C, {}^{n+1} \mathbf C_{\text{cr}},
{}^{n+1} \mathbf C_{\text{ii}}, {}^{n} \omega) \big]^{-1}  \ {}^{n} \mathbf C_{\text{cr}}.
\end{equation}
Let us consider the following fixed-point iteration for the solution of \eqref{ClassicalEBM1}
\begin{equation}\label{FixedPointIter}
{}^{n+1} \mathbf C_{\text{cr}}^{(0)} := {}^{n} \mathbf C_{\text{cr}}, \quad
{}^{n+1} \mathbf C_{\text{cr}}^{(i+1)} =
{}^{n} \mathbf C_{\text{cr}} + \Delta t \
\mathbf{f}_{\text{cr}}({}^{n+1} \mathbf C, {}^{n+1} \mathbf C_{\text{cr}}^{(i)},
{}^{n+1} \mathbf C_{\text{ii}}, {}^{n} \omega) \ {}^{n+1} \mathbf C_{\text{cr}}^{(i)}, \ i=0,1,2,...
\end{equation}
Since $\mathbf{f}_{\text{cr}}$ is a smooth function, the contractivity
condition is satisfied for sufficiently small $\Delta t$. Therefore,
the iterative process \eqref{FixedPointIter} converges to the exact solution of \eqref{ClassicalEBM1}.
To prove the symmetry of the solution ${}^{n+1} \mathbf C_{\text{cr}}$ pertaining to the classical EBM, it is sufficient to prove
that ${}^{n+1} \mathbf C_{\text{cr}}^{(i)} \in Sym$ for all $i=1,2,3,...$.
In other words, it suffice to prove that
\begin{equation}\label{WhatWeNeed}
\mathbf{f}_{\text{cr}}(\mathbf C, \mathbf C_{\text{cr}},
\mathbf C_{\text{ii}}, \omega) \ \mathbf C_{\text{cr}} \in Sym \quad
\text{for all} \quad \mathbf C, \mathbf C_{\text{cr}}, \mathbf C_{\text{ii}} \in Sym, \ \omega \in [0,1].
\end{equation}
Recall that
\begin{equation}\label{JustRecall}
\mathbf{f}_{\text{cr}}(\mathbf C, \mathbf C_{\text{cr}},
\mathbf C_{\text{ii}}, \omega) \ \mathbf C_{\text{cr}} =
2 \ \lambda (\sigma^{\lambda}_{\text{eq}},\omega)
\Big( \frac{\partial \sigma_{\text{eq}}}{\partial \tilde{\mathbf \Sigma}} \Big)^{\text{T}} \ \mathbf C_{\text{cr}}.
\end{equation}
In order to prove \eqref{WhatWeNeed}, we introduce a fictitious
deformation gradient and its parts as follows
\begin{equation}\label{FictitiousF}
\mathbf F = \mathbf C^{1/2}, \ \mathbf F_{\text{cr}} = \mathbf C_{\text{cr}}^{1/2}, \
\mathbf F_{\text{ii}} = \mathbf C_{\text{ii}}^{1/2}.
\end{equation}
For these artificially introduced quantities, all the relations
from Section 2.2 are valid.
In particular, a symmetric tensor $\hat{\mathbf \Sigma}$ exists such that
the right-hand side of \eqref{JustRecall}
is obtained from the symmetric tensor
$\lambda (\sigma^{\lambda}_{\text{eq}},\omega)
\frac{\partial \sigma_{\text{eq}}}{\partial \hat{\mathbf \Sigma}}$ by
the pull-back transformation  $ (\cdot) \mapsto \mathbf F^{\text{T}}_{\text{cr}} (\cdot) \mathbf F_{\text{cr}}$.
Since the pull-back and its inverse preserve the symmetry, the symmetry condition \eqref{WhatWeNeed} holds true.

An alternative (but more tedious) way of proving \eqref{WhatWeNeed}
is to note that $\mathbf C \tilde{\mathbf T}$
and $\mathbf C_{\text{cr}} \tilde{\mathbf X}$ are isotropic functions of $\mathbf C  \mathbf C^{-1}_{\text{cr}}$
and $\mathbf C_{\text{cr}}  \mathbf C^{-1}_{\text{ii}}$, respectively (cf. \cite{ShutovKreissig2008}),
and $\Big( \frac{\partial \sigma_{\text{eq}}}{\partial \tilde{\mathbf \Sigma}} \Big)^{\text{T}} $
is an isotropic function of $\tilde{\mathbf \Sigma} = \mathbf C \tilde{\mathbf T} - \mathbf C_{\text{cr}} \tilde{\mathbf X}$.

Now let us prove that the modified Euler-backward method \eqref{ModifEBMCcr} preserves the symmetry as well.
First we note that the modified method \eqref{ModifEBMCcr} yields the same solution as the
classical method \eqref{ClassicalEBM2} whenever ${}^{n} \mathbf C_{\text{cr}}$ is properly scaled in \eqref{ClassicalEBM2}:
\begin{multline}\label{Equivalence}
{}^{n+1} \mathbf C_{\text{cr}} = \overline{ \big[ \mathbf{1} - \Delta t \
\mathbf{f}_{\text{cr}}({}^{n+1} \mathbf C, {}^{n+1} \mathbf C_{\text{cr}},
{}^{n+1} \mathbf C_{\text{ii}}, {}^{n} \omega) \big]^{-1}}  \ {}^{n} \mathbf C_{\text{cr}}  \quad \Rightarrow   \\
\quad \text{there is} \ \beta>0 \ \text{such that} \quad
{}^{n+1} \mathbf C_{\text{cr}} =  \big[ \mathbf{1} - \Delta t \
\mathbf{f}_{\text{cr}}({}^{n+1} \mathbf C, {}^{n+1} \mathbf C_{\text{cr}},
{}^{n+1} \mathbf C_{\text{ii}}, {}^{n} \omega) \big]^{-1}  \ (\beta {}^{n} \mathbf C_{\text{cr}}).
\end{multline}
In other words, the modified EBM \eqref{ModifEBMCcr} is the classical EBM \eqref{ClassicalEBM2} where
the input quantity ${}^{n} \mathbf C_{\text{cr}}$ is scaled to enforce the
incompressibility relation $\det ({}^{n+1} \mathbf C_{\text{cr}})=1$.
Since the classical EBM \eqref{ClassicalEBM2} preserves the symmetry, so does its modification \eqref{ModifEBMCcr}.

\end{document}